\documentclass[%
 aip,pof,amsmath,amssymb,reprint,%
]{revtex4-1}

\usepackage{graphicx}
\usepackage{dcolumn}
\usepackage{bm}
\usepackage{color}
\newcommand{\smt}[1]{\textcolor{black}{#1}}
\newcommand{\jbm}[1]{\textcolor{black}{#1}}

\begin{document}


\title[DSS of Jets to Vortices]{\smt{Direct Statistical Simulation of Jets and Vortices in 2D Flows}}

\author{S. M. Tobias}
 \email{S.M.Tobias@leeds.ac.uk}
\affiliation{ 
Department of Applied Mathematics, University of Leeds. Leeds, LS2 9JT, UK
}%

\author{J. B. Marston}
\email{marston@brown.edu}
\affiliation{Department of Physics, Box 1843, Brown University, Providence, RI 02912-1843 USA}

\date{\today}

\begin{abstract}
In this paper we perform Direct Statistical Simulations \jbm{of a model of two-dimensional flow that exhibits a transition from jets to vortices.  The model employs} two-scale Kolmogorov forcing, \smt{with energy injected directly into the zonal mean of the flow}. We compare these results with those from  Direct Numerical Simulations. For square domains the solution takes the form of jets, but as the aspect ratio is increased a transition to isolated coherent vortices is found. We find that a truncation at second order in the equal-time but nonlocal cumulants that employs zonal averaging (zonal CE2) is capable of capturing the form of the jets for a range of Reynolds numbers as well as the transition to the vortex state, but, unsurprisingly, is unable to reproduce the correlations found for the fully nonlinear (\jbm{non-zonally} symmetric) vortex state. This result continues the program of promising advances in statistical theories of turbulence championed by Kraichnan. 

\end{abstract}

\maketitle

\section{\label{sec:Kraichnan}Direct Statistical Simulation: The legacy of Kraichnan}


Robert Kraichnan's vision of a statistical mechanics of turbulence notably emphasized essential differences between flows in two and three spatial dimensions \cite{Kraichnan:1967jk}.  Two dimensional flows are striking for the frequent emergence of coherent structures.  The structures are of two basic types:  Vortices and jets\cite{Kraichnan:1980uy,cho1996emergence}.  The Juno mission to Jupiter has recently returned beautiful images of both types of structures, with jets dominating at low latitudes, and a proliferation of vortices near the poles \cite{Bolton:2017ie}.  In this paper we investigate a simple model of two dimensional fluid flow that exhibits a transition between jets and vortices.  We employ both Direct \smt{Numerical} Simulation (DNS) and Direct Statistical Simulation (DSS).  DSS is a rapidly developing set of tools that attempt to describe, directly, the statistics of turbulent flows, bypassing the traditional way of accumulation \smt{of those statistics (for example mean flows and two-point correlation functions)} by DNS.  \smt{These statistical methods lead to a deeper understanding of fluid flows that should guide researchers to regimes not accessible through DNS.}

The pioneering work of Kraichnan largely focused on flows with isotropic and homogeneous statistics \cite{Herring1972}.  The statistical description of forward and/or inverse cascades of energy between different scales, \jbm{the topic explored in his seminal 1967 paper \cite{Kraichnan:1967jk}, is particularly clear in this context.}  Equally important, translational and rotational symmetries reduce the technical complexity of statistical theories.  Most fluid flows in nature, however, are both anisotropic and heterogeneous.   In DSS this is seen as a feature, rather than a defect, as anisotropy and inhomogeneity can lessen nonlinearity of the flows and make the statistics accessible to perturbative computation.  We show that a particularly simple version of DSS, one in which the equations of motion for the spatially-averaged statistics are closed at the level of second-order moments or cumulants \cite{marston65conover,tobias2011astrophysical}, is already able to reproduce many features of the model two-dimensional flow.  The Kolmogorov forcing we employ is purely deterministic and no parameters are tuned at the level of the second-order cumulant expansion (CE2), permitting a fair and unbiased comparison between DNS and DSS.  

We introduce the model in Section \ref{sec:level2}.  Results from DNS and DSS are presented in Section \ref{sec:results}.  Comparison between the two approaches is made in Section \ref{sec:comparison} and some conclusions and possible directions for further exploration are discussed in Section \ref{sec:conclusion}. 

\section{\label{sec:level2}Set-up of the model: formulation, equations, and forcing.}

The models we study are of incompressible fluid moving \smt{on a two-torus --- i.e\ a  two
dimensional Cartesian domain ($0 \le x < L_x$, $0 \le y < L_y = 2 \pi$) with periodic boundary conditions \jbm{in both directions}}.  The fluid motion is damped by viscosity $\nu$ and driven
by a time-independent forcing (described below).  Owing to the two-dimensionality of the system the dynamics is completely described by 
the time-evolution of the vorticity $\zeta \equiv \hat{z}
\cdot (\vec{\nabla} \times \vec{v})$, which  (in dimensional units) is given by  
\begin{eqnarray}
\dot{\zeta} + J(\psi,\zeta) = \nu \nabla^2 \zeta + g(y),
\label{EOM}
\end{eqnarray}
where $g(y)$ is the forcing term and $J(A,B)$ is the Jacobian operator given by $J(A,B)= A_x B_y - A_y B_x$.
We note that, in contrast to some earlier models, here we do not consider the effects of rotation via a $\beta$-effect.

The forcing $g(y)$ is a generalisation of the Kolmogorov forcing (see e.g. \onlinecite{lk2014}) to two \smt{meridional} wavenumbers; that is we set
\begin{eqnarray}
g(y)= A_1 \cos(y)+ 4.0*A_4 \cos (4y).  
\label{forcingdep}
\end{eqnarray}
We set $A_1=-1$ and $A_4=-2$, which leads to non-trivial dynamics in the fluid system. \smt{We note here that this choice of deterministic forcing injects energy directly into the zonal flow, and should be contrasted with previous studies that impose stochastic forcing in the small zonal scales.}
Once the forcing and the length scale in $y$ (say) is fixed then the dynamics (and indeed statistics) of
the flow is determined by the choice of the viscosity $\nu$ (which controls the Reynolds number of the flow, which may be calculated {\it a posteriori}) and the
aspect ratio (determined by $L_x$).

The aim of this paper is to determine how successful Direct Statistical Simulation truncated at second order (sometimes termed CE2) is at describing the transitions that occur as the parameters are varied. In particular we shall investigate how well DSS reproduces the strength of the mean shear flows and the transition from solutions dominated by jets to those dominated by coherent vortex pairs.


\section{Results}
\label{sec:results}

In this section we \smt{describe} results from DNS (in subsection \ref{dns}) and those obtained from DSS using CE2 (in subsection \ref{dss}) before \smt{comparing them in Section IV}.

\subsection{\label{dns} Direct Numerical Simulation}

\begin{figure}
\centerline{\includegraphics[width=2.5in]{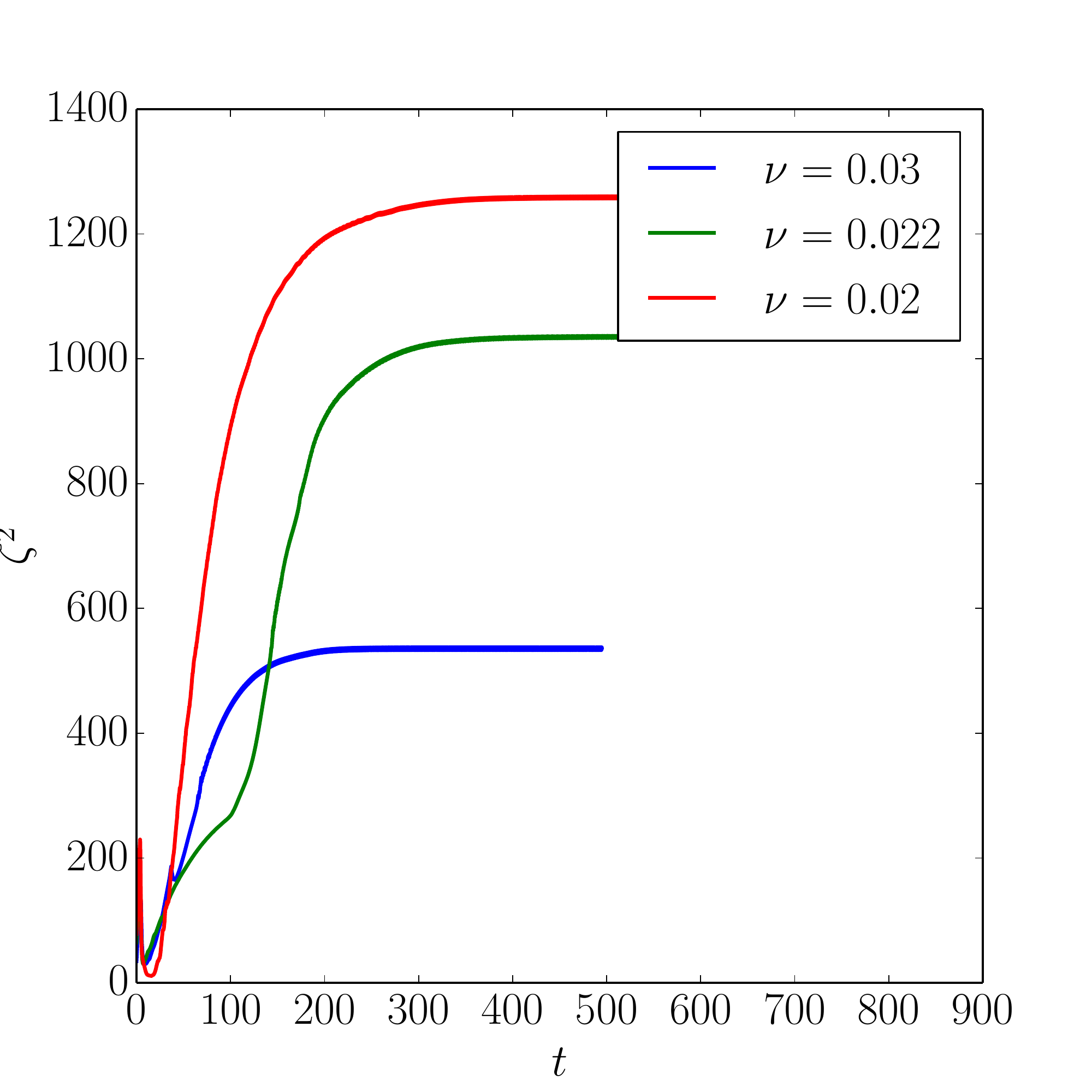}}
\centerline{\includegraphics[width=2.5in]{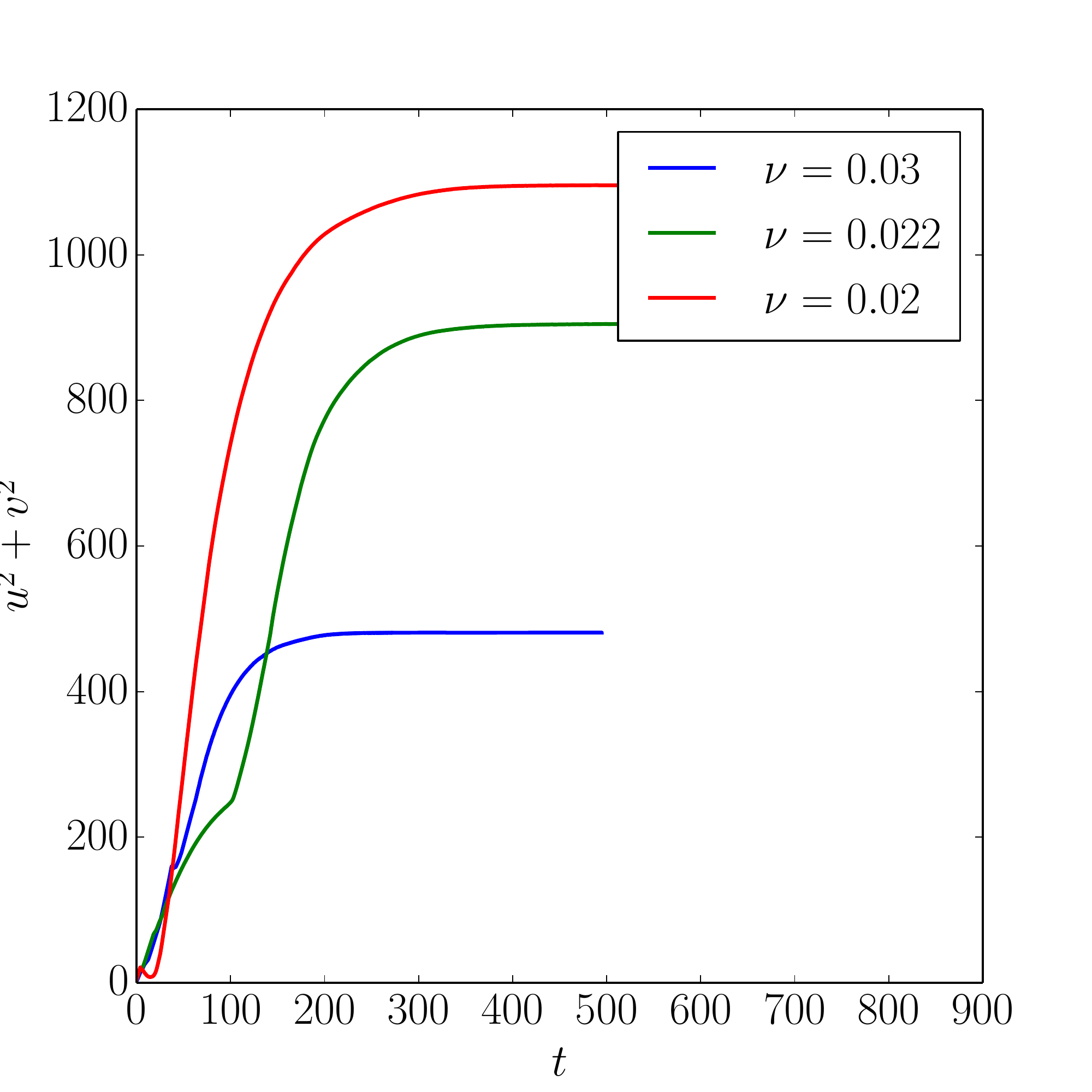}}
\caption{Enstrophy and Kinetic Energy timeseries for the solutions at three different viscosities for $L_x = L_y = 2 \pi$.  Both of these increase with decreasing viscosity.}
\label{fig:ensen}
\end{figure}

Direct Numerical simulation is performed using a pseudo-spectral code optimised for use on parallel architectures with typical resolutions of $512^2$. In all cases the resolution is increased \smt{to this level to obtain until convergent results.}
Initially we integrate the equations for  three different values of the viscosity in a square domain with
$L_x=L_y=2\pi $. The time series for the resulting spatially-averaged enstrophy $\overline{\zeta^2}$ and kinetic energy $\overline{\psi_y^2+\psi_x^2}$ where
\begin{equation}
\overline{A}  \equiv \frac{1}{L_x L_y}\iint A \,dx \,dy,
\end{equation} 
are shown in Figure~\ref{fig:ensen} for three values of the viscosity.
This figure clearly shows that, as expected, as the viscosity is decreased (with the forcing fixed) both the enstrophy (top panel) and kinetic energy (bottom panel) of the solutions increases. 

\begin{figure*}
\centerline{\includegraphics[width=2.0in]{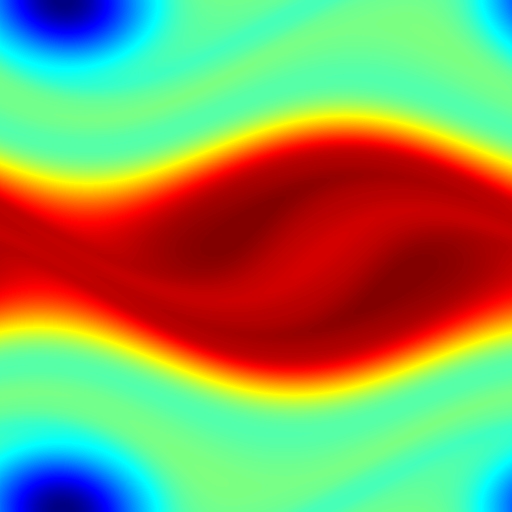} \quad \includegraphics[width=2.0in]{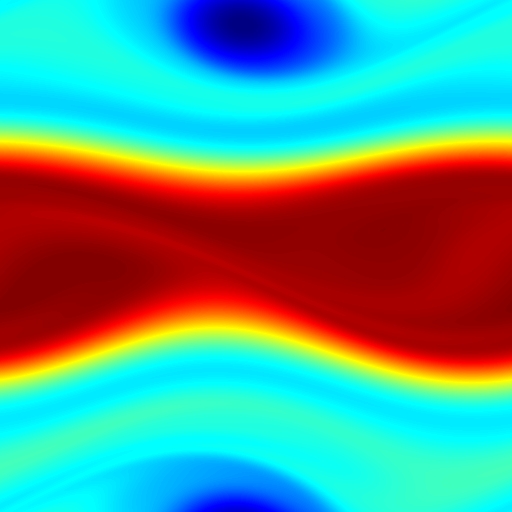} \quad 
\includegraphics[width=2.0in]{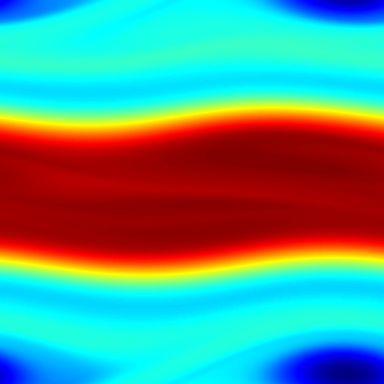}}
\caption{Snapshots of vorticity for (a) $\nu =0.03$ (b) $\nu = 0.022$ (c) $\nu=0.02$, with $L_x=L_y = 2 \pi$. For these figures the colours are scaled between (a) $[-61,34]$,  (b) $[-67,46]$, (c) $[-75,52]$. Movies showing the dynamics for these parameters are contained in the supplementary material (multimedia view).}
\label{fig:snapshots}
\end{figure*}

Figure~\ref{fig:snapshots} (multimedia view) shows snapshots from movies of the evolution of the vorticity for the cases with $L_x=2 \pi$ (which are included in the supplementary material). 
The flow is reasonably laminar although the solution has already undergone a bifurcation from a steady state. After some initial transients the solution becomes time periodic, with a strong band/jet of positive vorticity in the domain and weaker negative vorticity (in the form of a vortex) at the edges. This corresponds to a rightward jet in the upper half of the domain and a reverse jet in the lower half (see later). Both the jet and the vortex remain fixed in space though pulse in time. Though time-dependence is present, these vortex regions possess a well defined zonal  mean, which can be calculated by averaging over a suitably long time. The average Reynolds number for this flow is given by $Re \approx 730$.

As the viscosity is decreased from this solution the dynamics becomes more irregular and time-dependent, as shown in the  two movies.
Decrease of the viscosity leads to stronger patches of vorticity and faster flows. For $\nu = 0.022$, $Re \approx 1370$, whilst for $\nu=0.02$, $Re \approx 1650$.  For both of these solutions the \jbm{non-zonally} symmetric part   \smt{i.e. the part of the solution with $k_x \ne 0$} and the vortex travel in space, rather than remaining fixed as for the earlier case. The strength of the vortex patches and jets increases with decreasing $\nu$ (as does the corresponding mean flows and vorticities --- see later) as the inertial terms play an increasingly important role. 

\begin{figure}
\centerline{\includegraphics[width=2.5in]{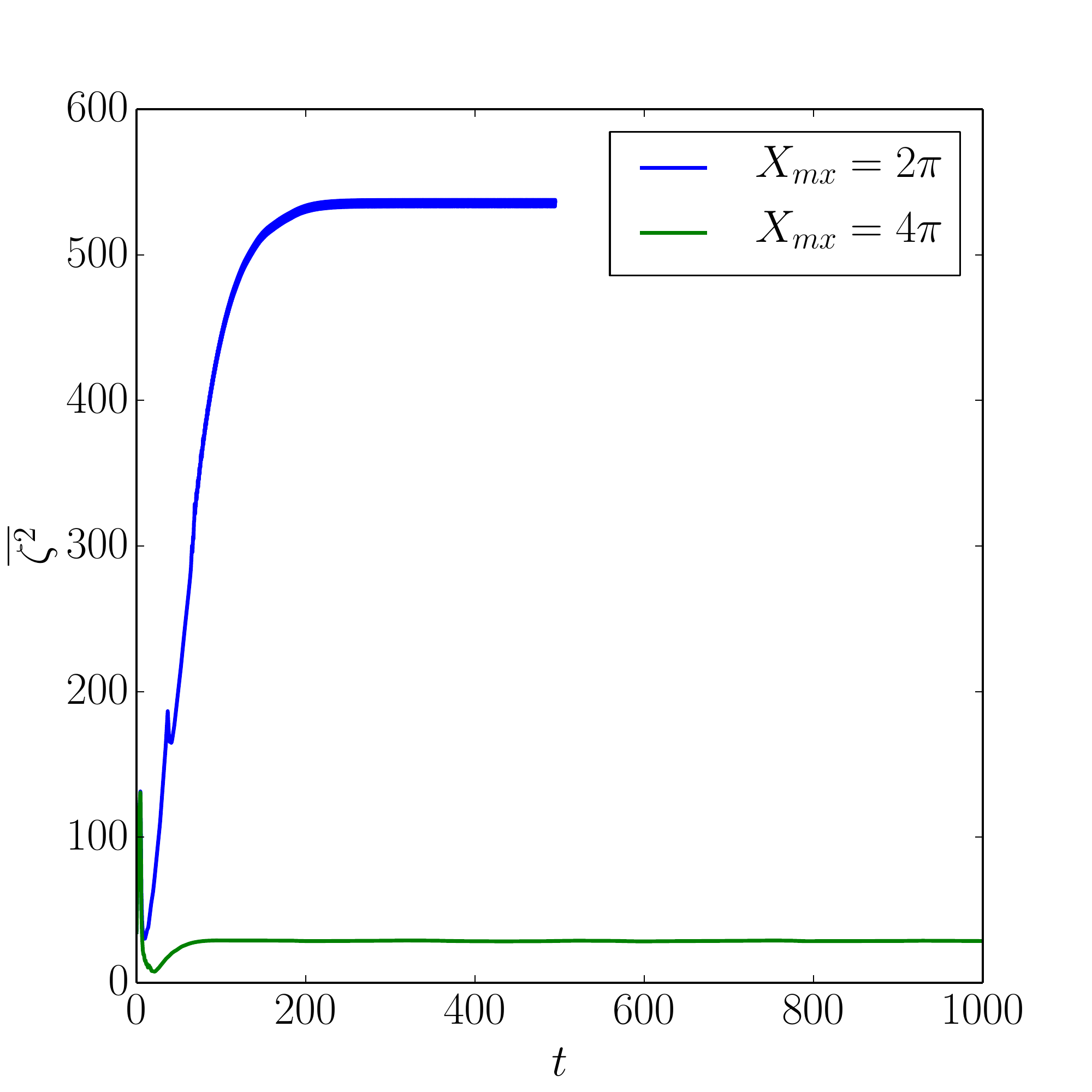}}
\centerline{\includegraphics[width=2.5in]{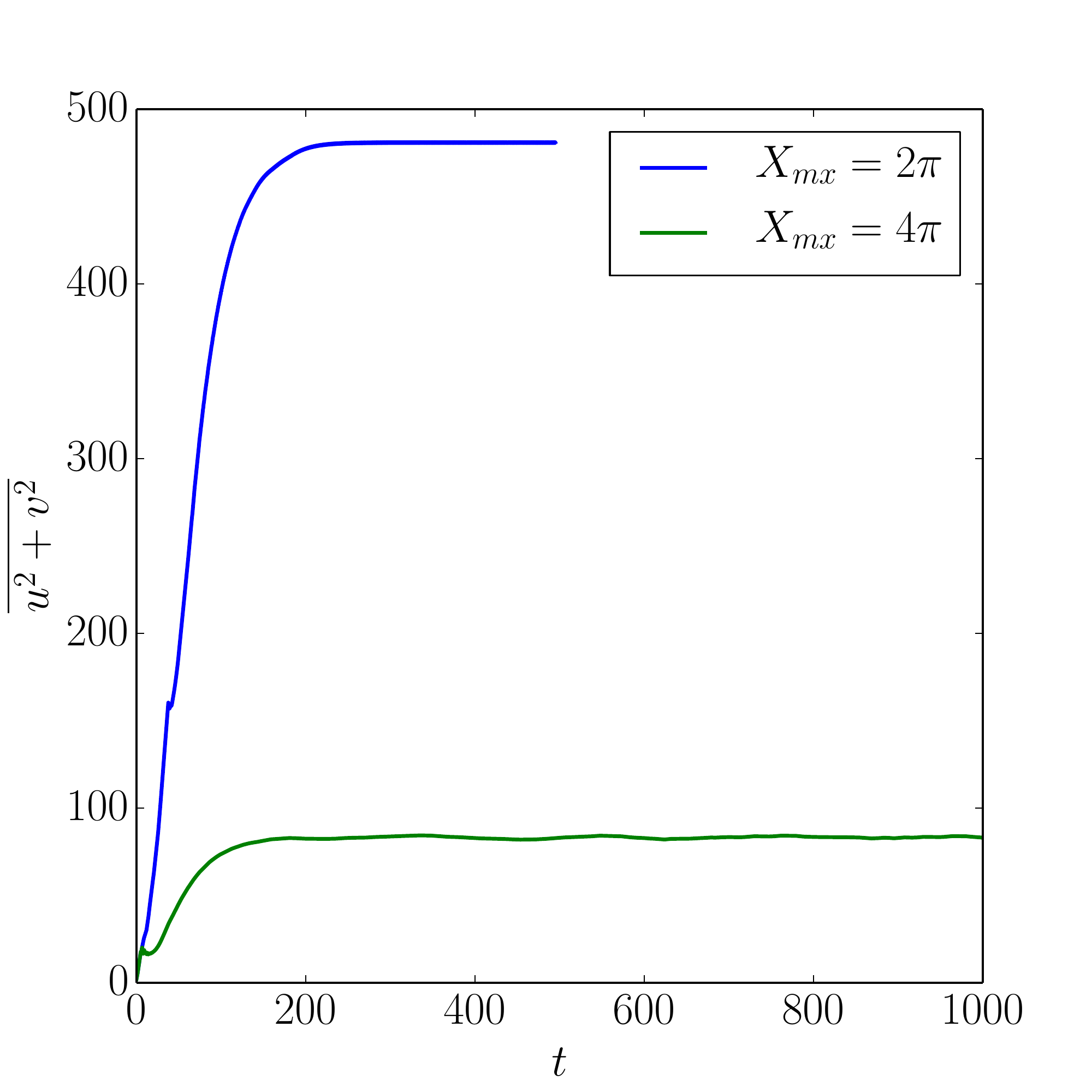}}
\caption{Enstrophy and Kinetic Energy timeseries for the solutions at $\nu = 0.03$ and  $L_x = 2 \pi$ and $L_x = 4 \pi$.  The solutions for the larger aspect ratio are significantly weaker in energy and have less enstrophy.}
\label{fig:vortex_en_ens}
\end{figure}

\begin{figure*}
\centerline{\includegraphics[width=5in]{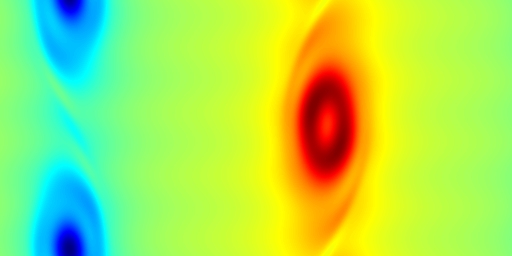}}
\caption{Snapshot of vorticity for $\nu =0.03$ (b) $L_x=4 \pi$ and $L_y = 2 \pi$.  For this figures the colours are scaled between $[-23,18]$. A movie showing the dynamics for these parameters is contained in the supplementary material (multimedia view).}
\label{fig:snapshot_vort}
\end{figure*}

When the aspect ratio is increased so that $L_x=4 \pi$, the nature of the solution changes. The driving which is independent of $x$ no longer puts substantial power into the $k_x=0$ modes, but instead drives a fully nonlinear quasi-steady vortex pair solution as shown in Figure~\ref{fig:snapshot_vort} (multimedia view) and the corresponding movie. This state is reminiscent of the localised states analysed extensively in Ref.~\onlinecite{lk2014}. For this state the average enstrophy and vorticity are significantly lower than for the jet states (as shown in the time series in Figure~\ref{fig:vortex_en_ens}). Moreover, as we shall see, this state has little energy in the zonally-averaged vorticity and flow and so can not be characterised as a jet state. This remarkable transition appears to be the opposite of a {\it zonostrophic instability} (see Ref.~\onlinecite{sy2012}); there a small-scale forcing with zero zonal mean drives flow that interacts with rotation to put significant amount of energy into a zonally averaged jet. Here the forcing is designed to drive strong zonal flows, but nonlinear interactions prefer to put energy into vortex states with weak zonal flows, and may therefore be termed a ``vortostrophic instability.''  We note that the aspect ratio controls a similar transition between jets and vortices in other two-dimensional models \cite{Bouchet:2009cl,Bouchet:2012ea,Laurie:2015co,Frishman:2017jy}. The non-trivial nonlinear dynamics provides an interesting testing ground for the types of statistical theories favoured by Kraichnan and so, in the next section, we compare the results obtained here via Direct Numerical Simulation, with those obtained by Direct Statistical Simulation truncated at second order (CE2).

\subsection{
\label{dss} Direct Statistical Simulation: The Cumulant Equations}

In this section we perform DSS for the system for the same range of parameters as above.  The approach we take is based upon truncating the hierarchy of equations of motion for the equal-time cumulants at low order.  It is related to stochastic structural stability theory (S3T) \cite{Farrell:2007fq,Constantinou:2013fh} and other approaches \cite{Laurie:2014dn,Falkovich:2016ih} that do not assume spatial homogeneity or isotropy in the statistics.  Here we define the cumulants in terms of zonal averages over the x-direction (see Refs. \onlinecite{marston65conover,tobias2011astrophysical,tobias2013direct,marston2014direct,ait2016cumulant}) as opposed to ensemble averages \cite{bakas2013emergence,Bakas:2015iy,altan2017}.  Thus
\begin{equation}
 c_{\zeta}(y) = \langle \zeta \rangle,
\end{equation}
where $\langle \rangle$ indicates a zonal average, is the first cumulant and 
\begin{equation}
 c_{\zeta \zeta}(y, y^\prime,\xi) = \langle \zeta^\prime(x, y) \,\zeta^\prime(x+\xi, y^\prime) \rangle,
\end{equation}
is the second cumulant (or two-point correlation function). We note that owing to the translational symmetry of the system (including the forcing) the first cumulant is a function only of $y$ and the second cumulant is a function of three rather than four dimensions \cite{tobias2013direct,marston2014direct}. There are similar definitions for the first and second cumulants involving the streamfunction (i.e. $c_\psi$ and $c_{\psi \zeta}$), but these can be related straightforwardly to the cumulants for the vorticity.

The cumulant hierarchy can be derived in a number of ways \cite{marston2014direct}. Truncated at second order (CE2) \smt{this} takes the form of evolution equations for $c_{\zeta \zeta}(y,y^\prime)$ and $c_\zeta(y)$ (see Ref. \onlinecite{tobias2013direct}):
\begin{eqnarray}
\frac{\partial }{\partial t}  c_{\zeta}(y) &=&
\left[ -\left(\frac{\partial}{\partial y}+ \frac{\partial}{\partial y^\prime}\right)\frac{\partial}{\partial \xi} c_{\psi \zeta(y,y^\prime,\xi)}\right] \Biggr |_{y^\prime=y, \xi = 0} \nonumber \\
&+&g(y)+\nu \frac{\partial^2 }{\partial y^2} c_\zeta(y),
\label{czeteqn}
\end{eqnarray}

together with

\begin{eqnarray}
\frac{\partial }{\partial t}  c_{\zeta \zeta} &=& 
\frac{\partial }{\partial y} c_\psi(y) \frac{\partial }{\partial \xi} c_{\zeta \zeta}(y,y^\prime,\xi) 
\nonumber \\
&-& \frac{\partial }{\partial y} (c_\zeta(y)) \frac{\partial }{\partial \xi} c_{\psi \zeta}(y,y^\prime,\xi)  
\nonumber \\
&-&\frac{\partial }{\partial y^\prime} c_\psi(y^\prime) \frac{\partial }{\partial \xi} c_{\zeta \zeta}(y,y^\prime,\xi)  \nonumber \\
&-& \frac{\partial }{\partial y^\prime} (c_\zeta(y^\prime)) \frac{\partial }{\partial \xi} c_{\zeta \psi}(y,y^\prime,\xi)  
\nonumber \\
&+&\nu \left( 2\frac{\partial^2}{\partial \xi^2}+\frac{\partial^2}{\partial y^2}+\frac{\partial^2}{\partial {y^\prime}^2}\right) c_{\zeta \zeta}.
\label{czetzeteqn}
\end{eqnarray}

In the limit of no forcing or dissipation, equations~(\ref{czeteqn}--\ref{czetzeteqn}) conserve energy, enstrophy and the Kelvin impulse; thus the CE2 is a conservative approximation \cite{marston65conover,marston2014direct}.  The equations are integrated forward in time numerically using a pseudospectral code with typical resolutions of $(n_y,n_{y^\prime},n_\xi) = (64,64,16)$ until the statistics have settled down to a statistically steady state and means and two-point correlation functions are averaged in time.

\section{Comparison of DNS and DSS}
\label{sec:comparison}

The solutions from the cumulant equations are compared with the corresponding statistics obtained from DNS by averaging in $x$ and time. In all cases the DNS solutions are averaged over the final third of the evolution and the statistics are well converged. The DSS solutions were averaged over the final $10\%$ of the evolution, though the averages rapidly converge in all cases.

\begin{figure*}
\centerline{\includegraphics[width=2.5in]{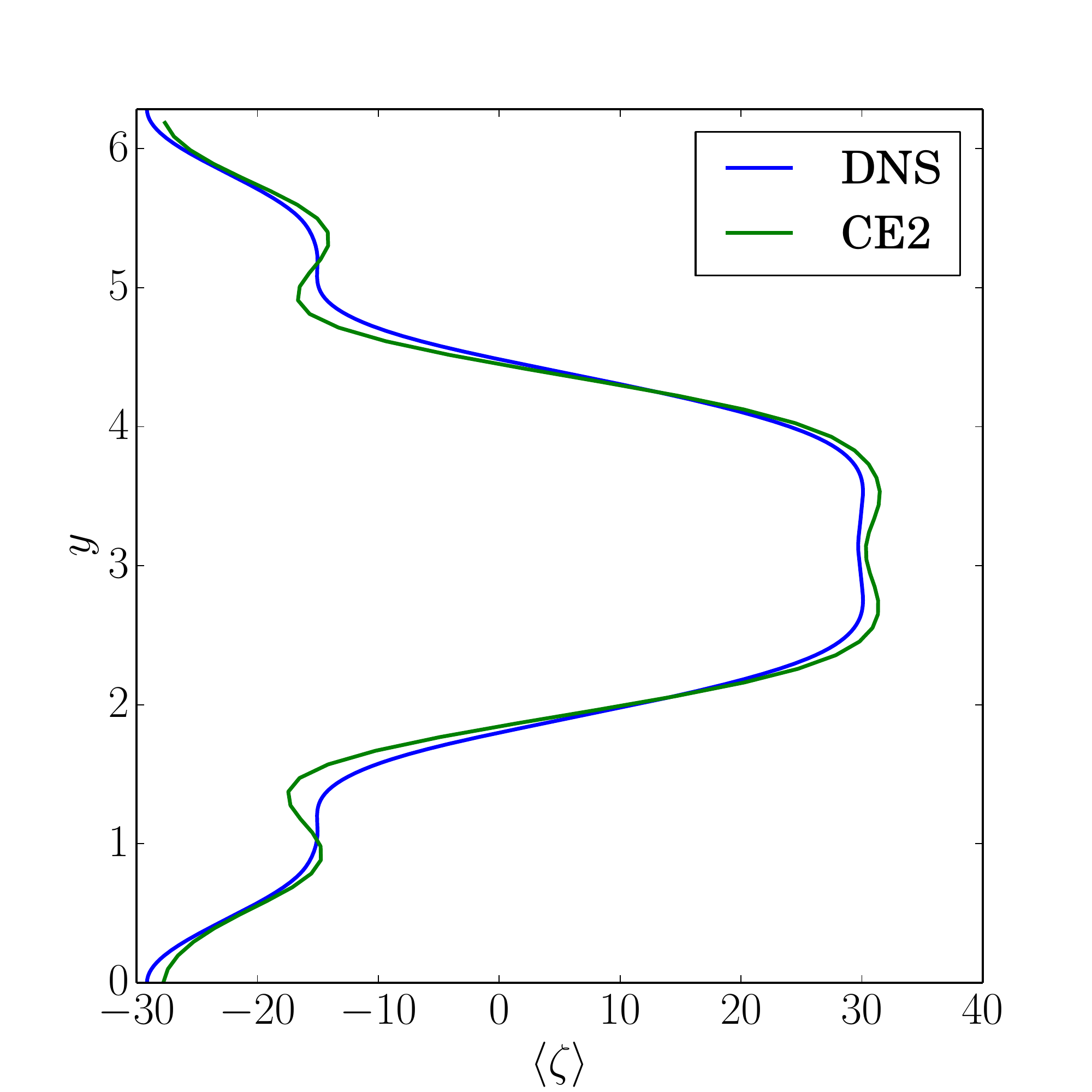}\quad\includegraphics[width=2.5in]{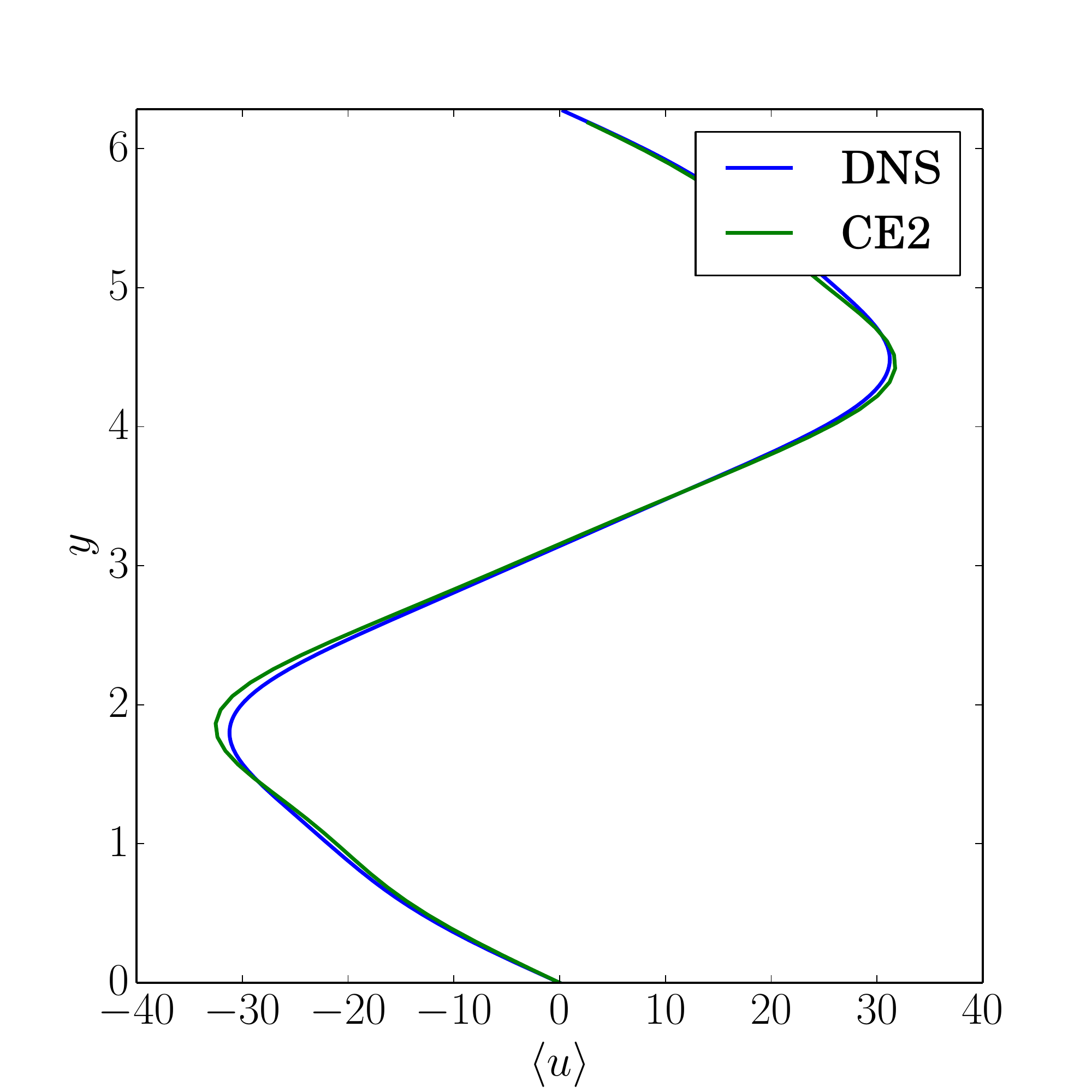}}
\centerline{\includegraphics[width=2.5in]{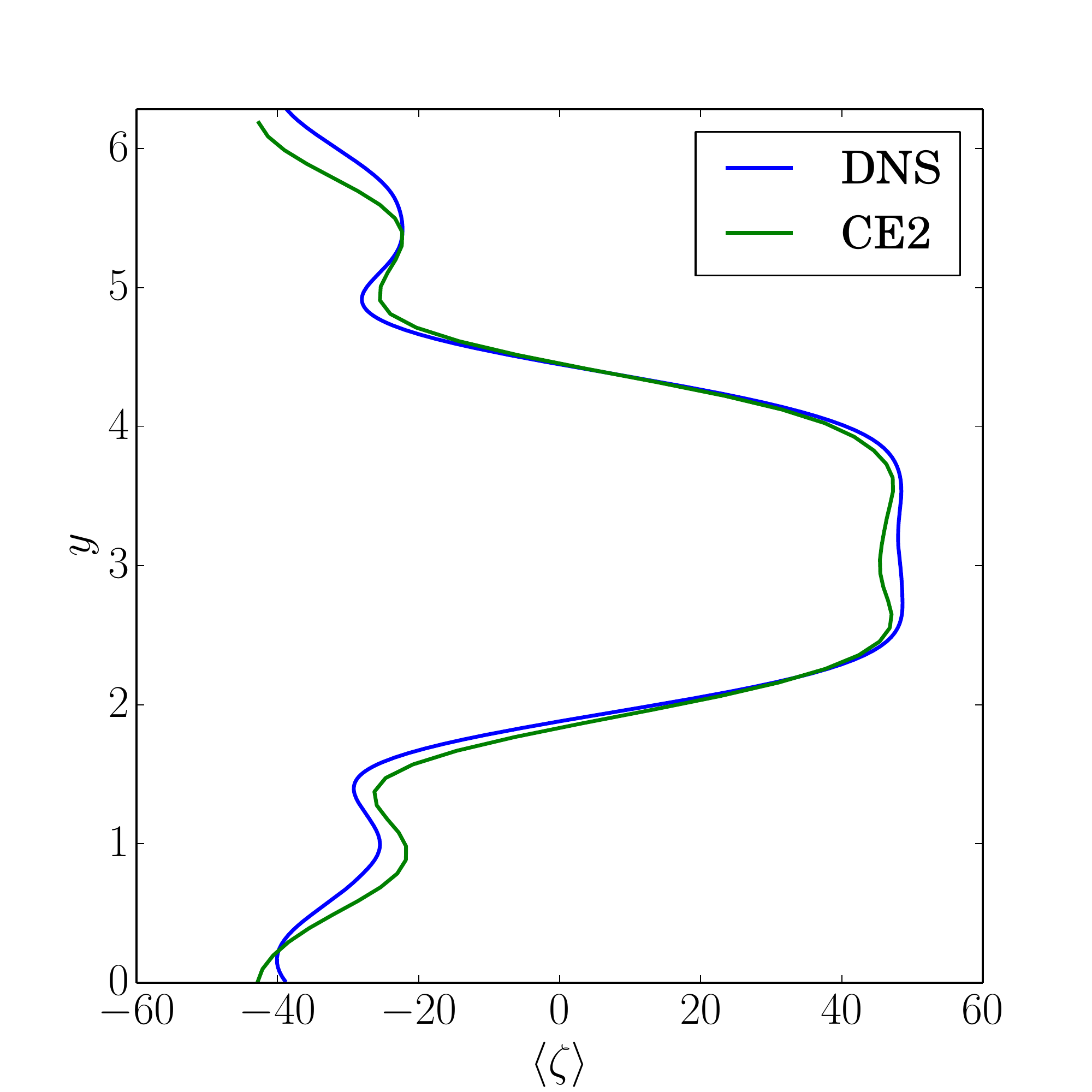}\quad\includegraphics[width=2.5in]{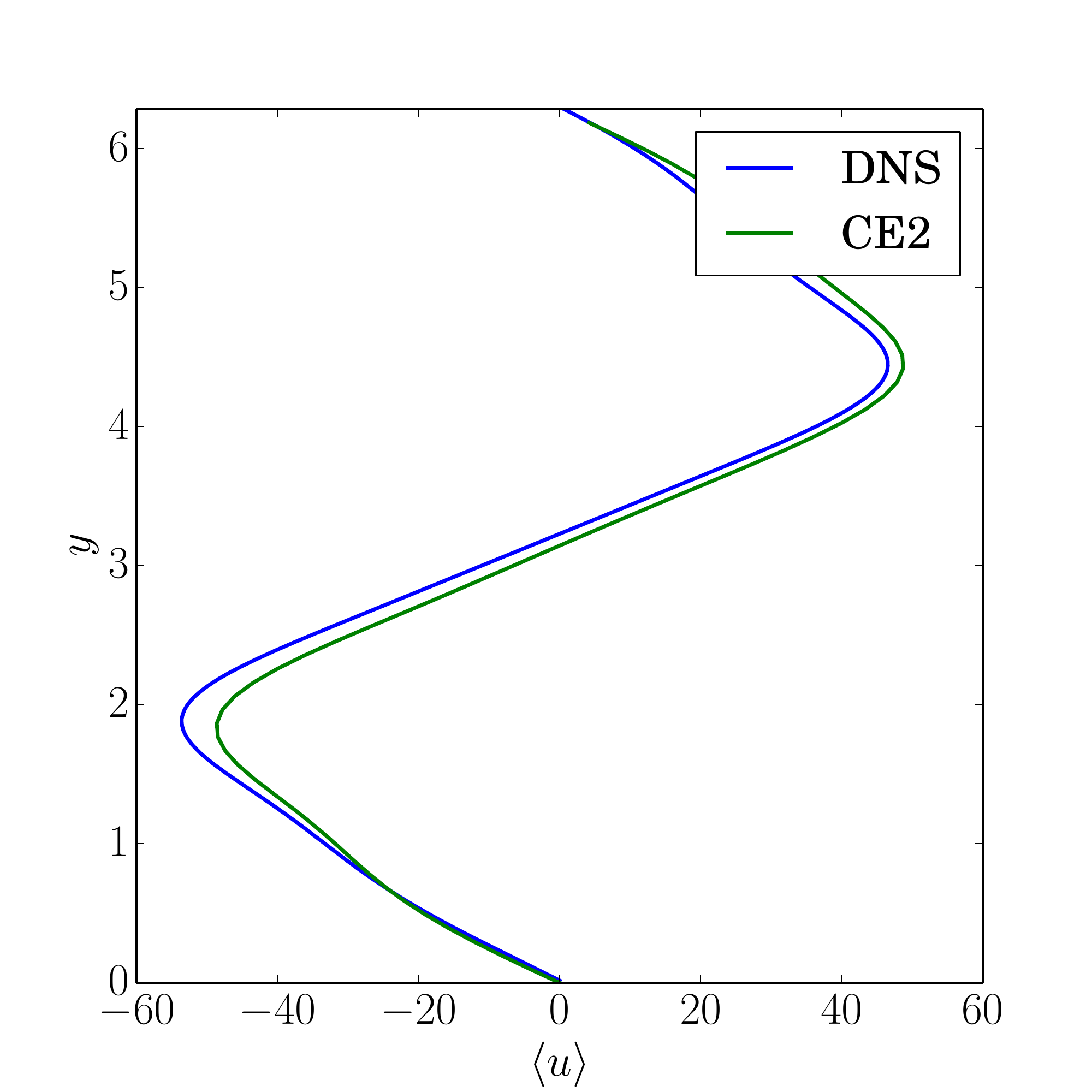}}
\centerline{\includegraphics[width=2.5in]{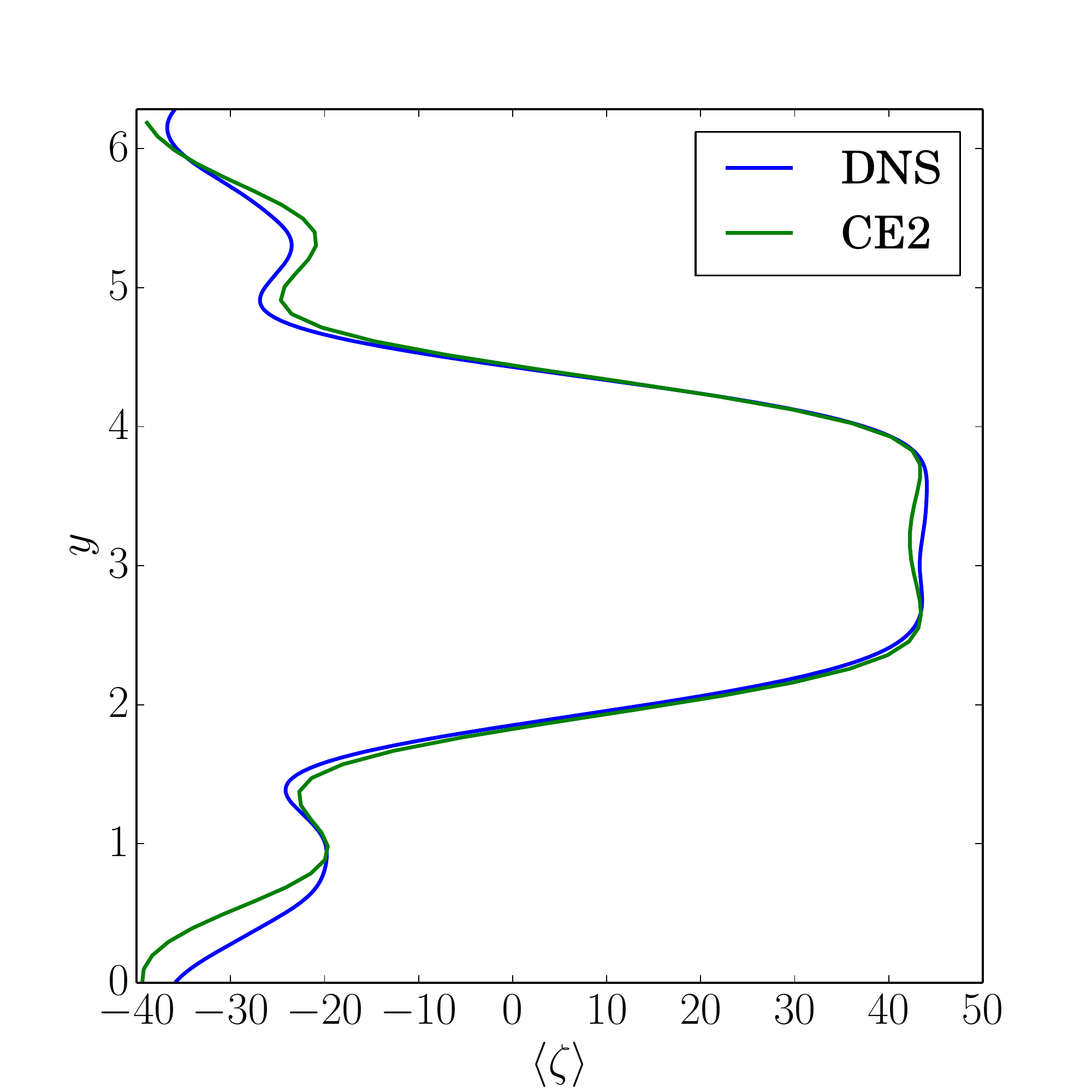}\quad\includegraphics[width=2.5in]{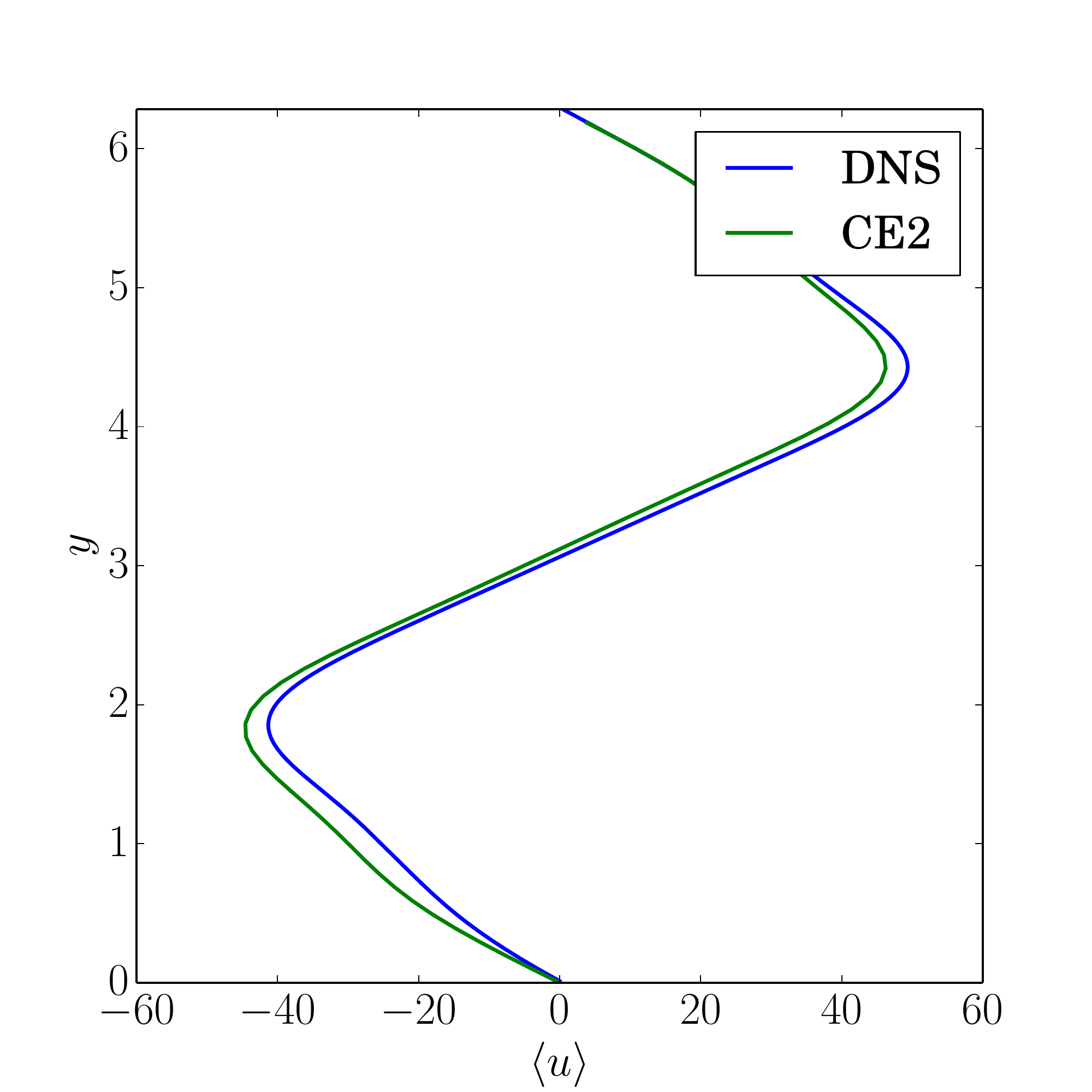}}
\caption{Comparison of DNS and CE2. Mean vorticity and zonal flow for $\nu = 0.03$ (top row) $\nu=0.022$ (middle row) $\nu =0.02$ (bottom row).}
\label{fig:means}
\end{figure*}  

Figure~\ref{fig:means} shows the comparison between DNS and CE2 for a domain of length $2 \pi$ and decreasing viscosity. As the viscosity is decreased the mean vorticity amplitude $\langle \zeta \rangle$ and the mean zonal flow $\langle u \rangle$ increase in amplitude as expected. It is clear that the comparison of the mean flows between CE2 and DNS is excellent. CE2 has a tendency to emphasise turning points in the vorticity that are washed out by eddy + eddy $\rightarrow$ eddy interactions in the DNS \jbm{\cite{marston65conover}}. However the agreement in the amplitude and form of the solution is very good. \smt{One might expect CE2 to improve as the ratio of energy in the zonal mean flow to that in the fluctuations increases. This expectation is largely met, though in all cases agreement is good.}

\begin{figure*}
\centerline{\includegraphics[width=2.5in]{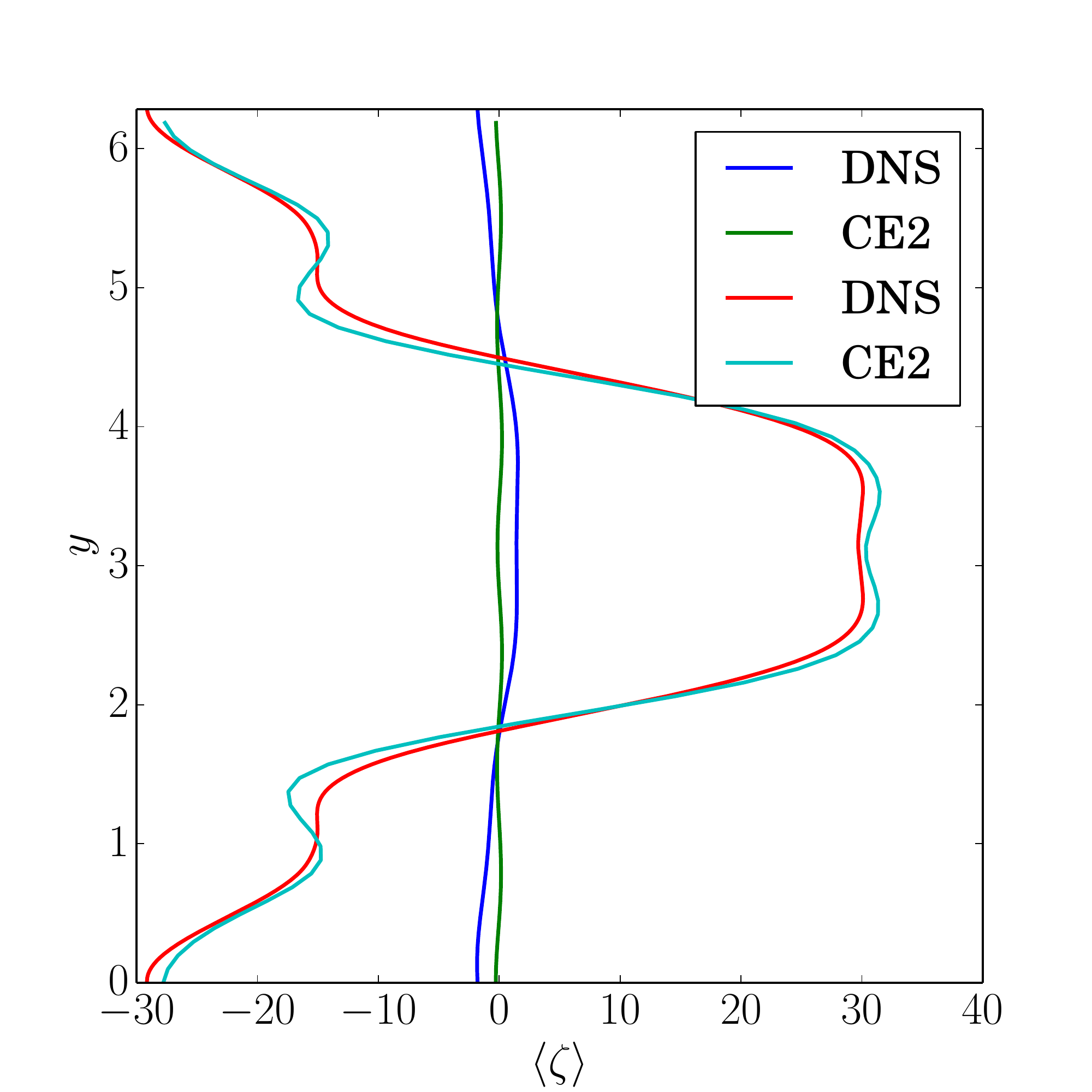}\quad\includegraphics[width=2.5in]{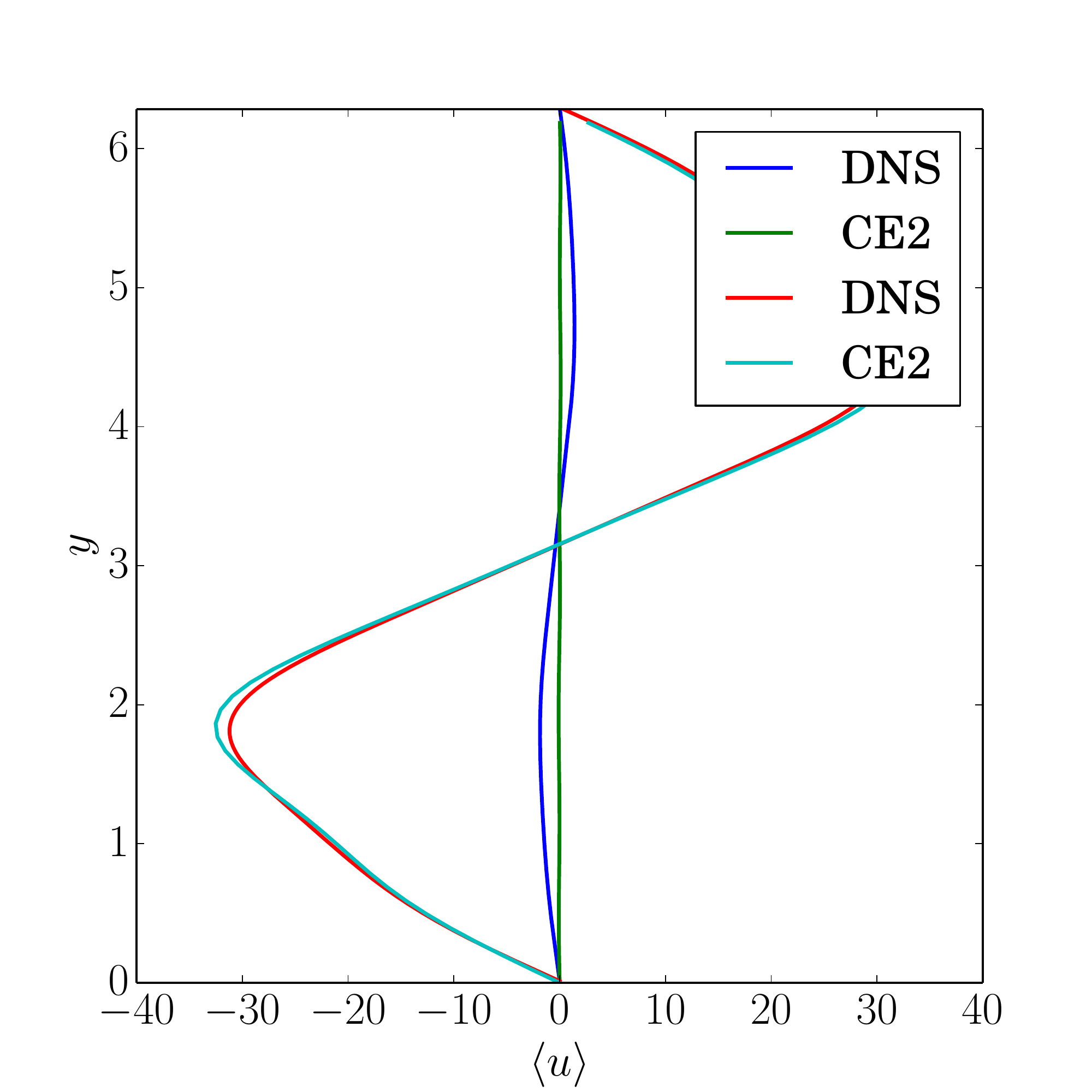}}
\caption{Comparison of DNS and CE2. Mean vorticity and zonal flow for $\nu = 0.03$ and $L_x = 2 \pi$ (red and cyan)  and  $L_x = 4 \pi$ (blue and green).}
\label{fig:trans_mean}
\end{figure*}

What is remarkable is that CE2 is capable of capturing the transition to vortices as shown in Figure~\ref{fig:trans_mean}. When the zonal flow is switched off in DNS by changing the aspect ratio, CE2 predicts the same behaviour. Although CE2 does not get the form of the weak zonal flow completely correct, it predicts the amplitude very well. This is unexpected since zonally averaged CE2 is expected to work poorly in a case where the zonal means are small and subdominant to the fluctuations.

\begin{figure*}
\centerline{\includegraphics[width=2.5in]{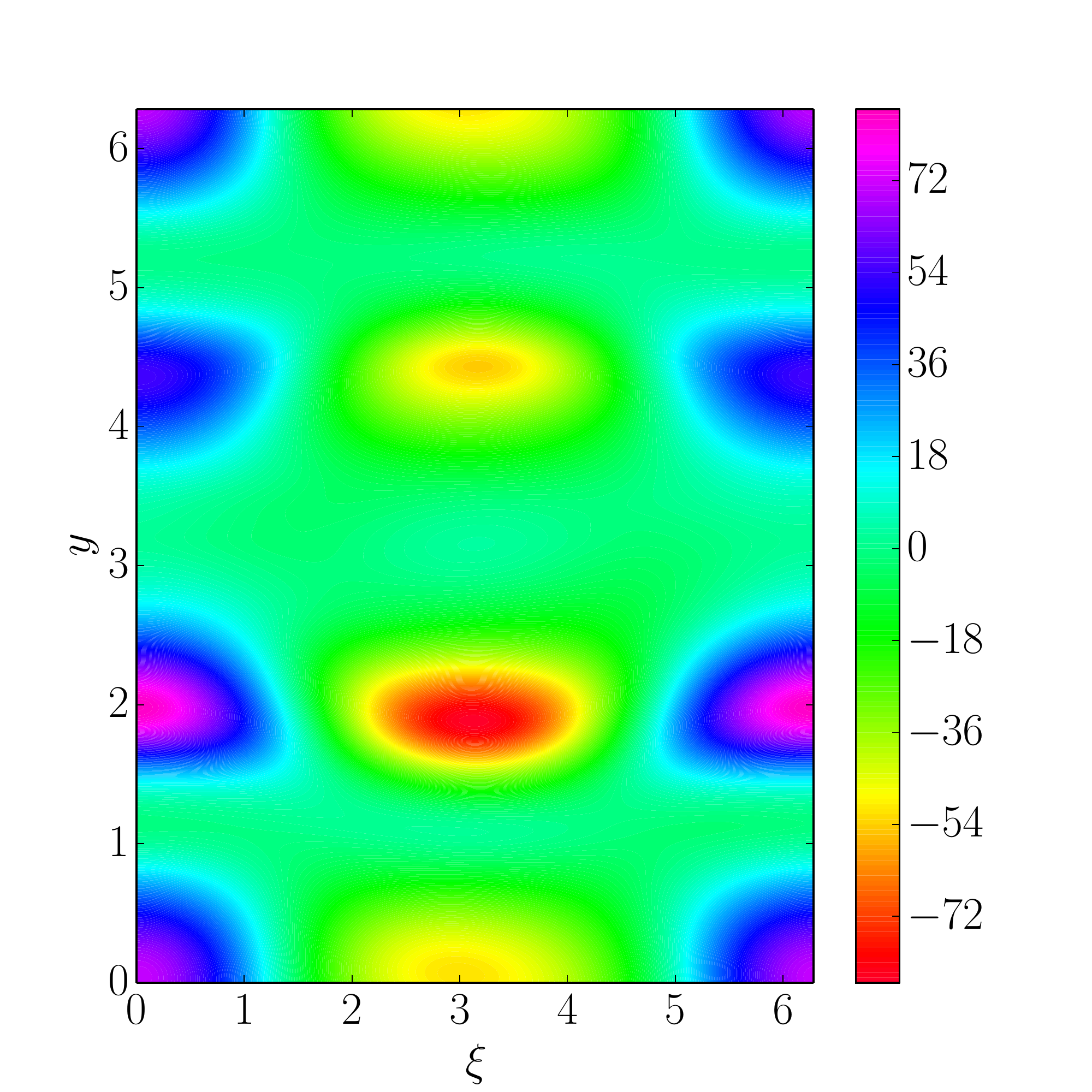} \quad \includegraphics[width=2.5in]{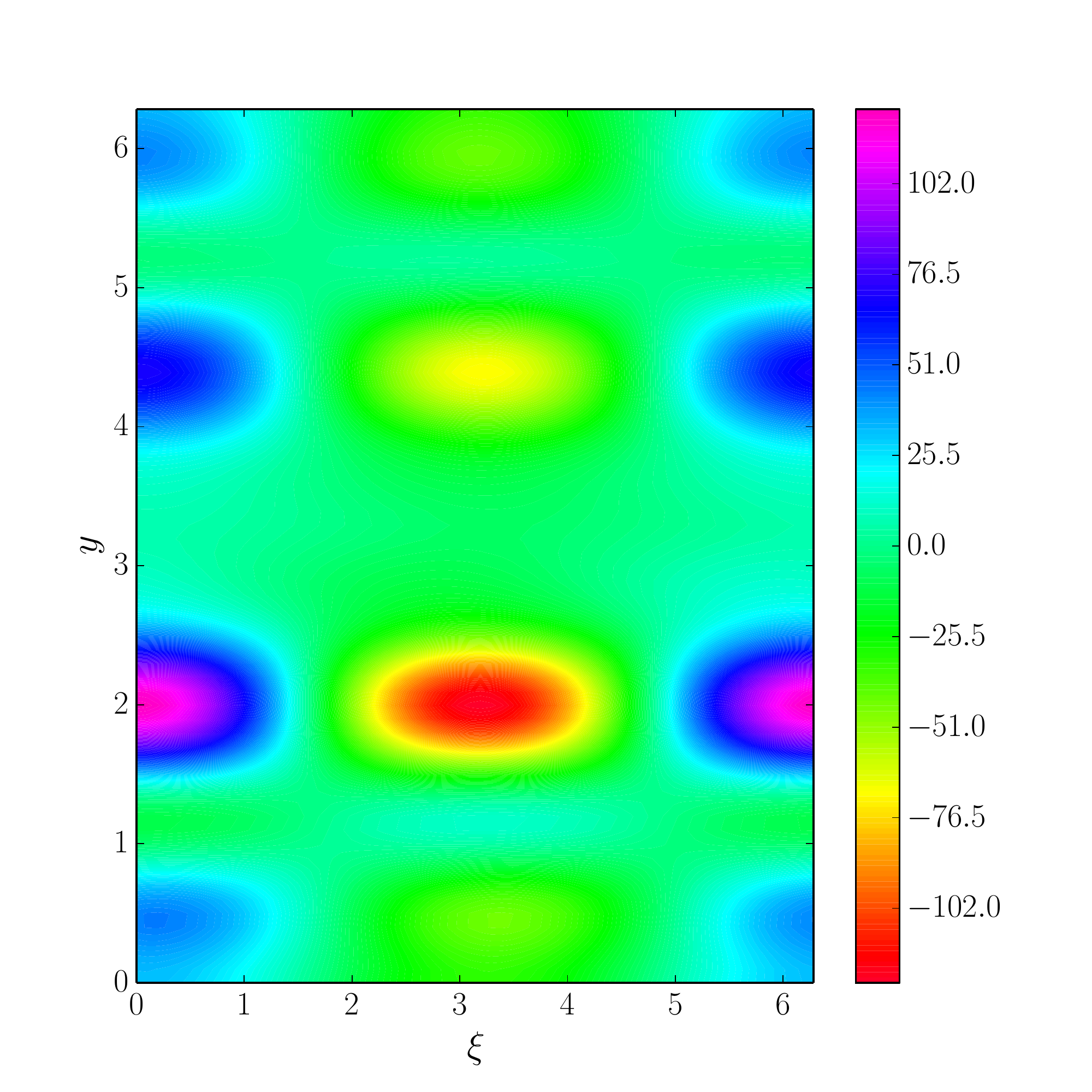}} 
\centerline{\includegraphics[width=2.5in]{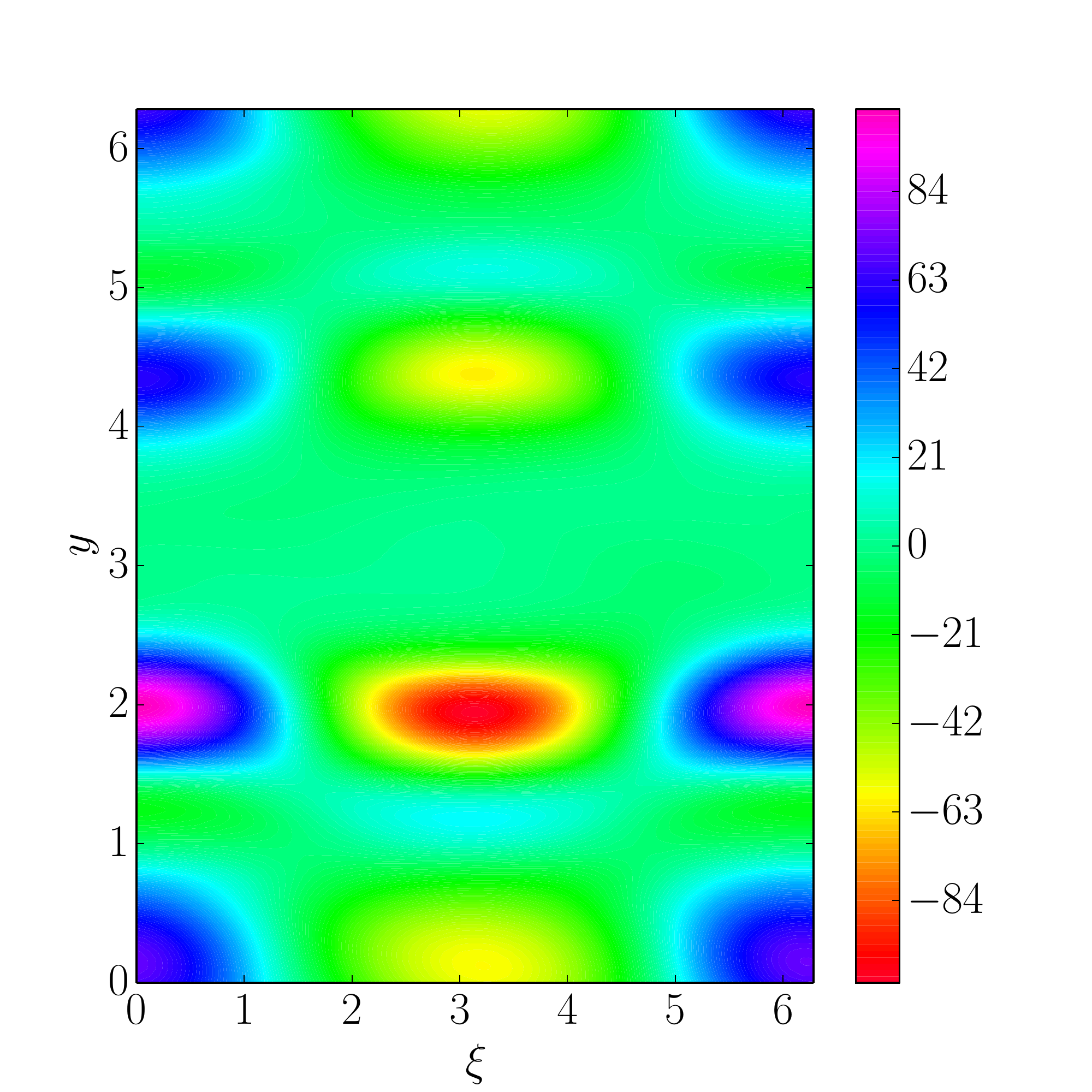} \quad \includegraphics[width=2.5in]{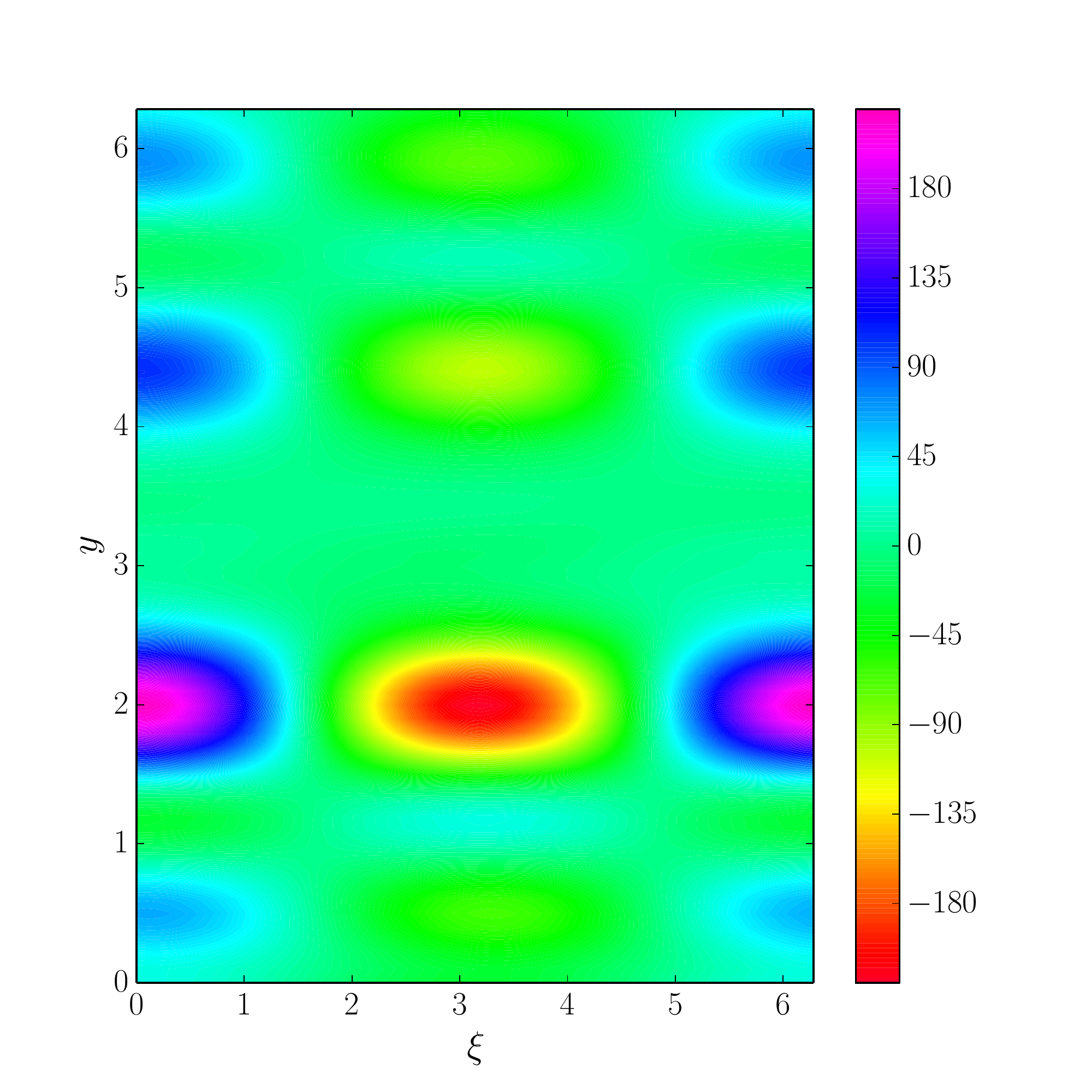}} 
\centerline{\includegraphics[width=2.5in]{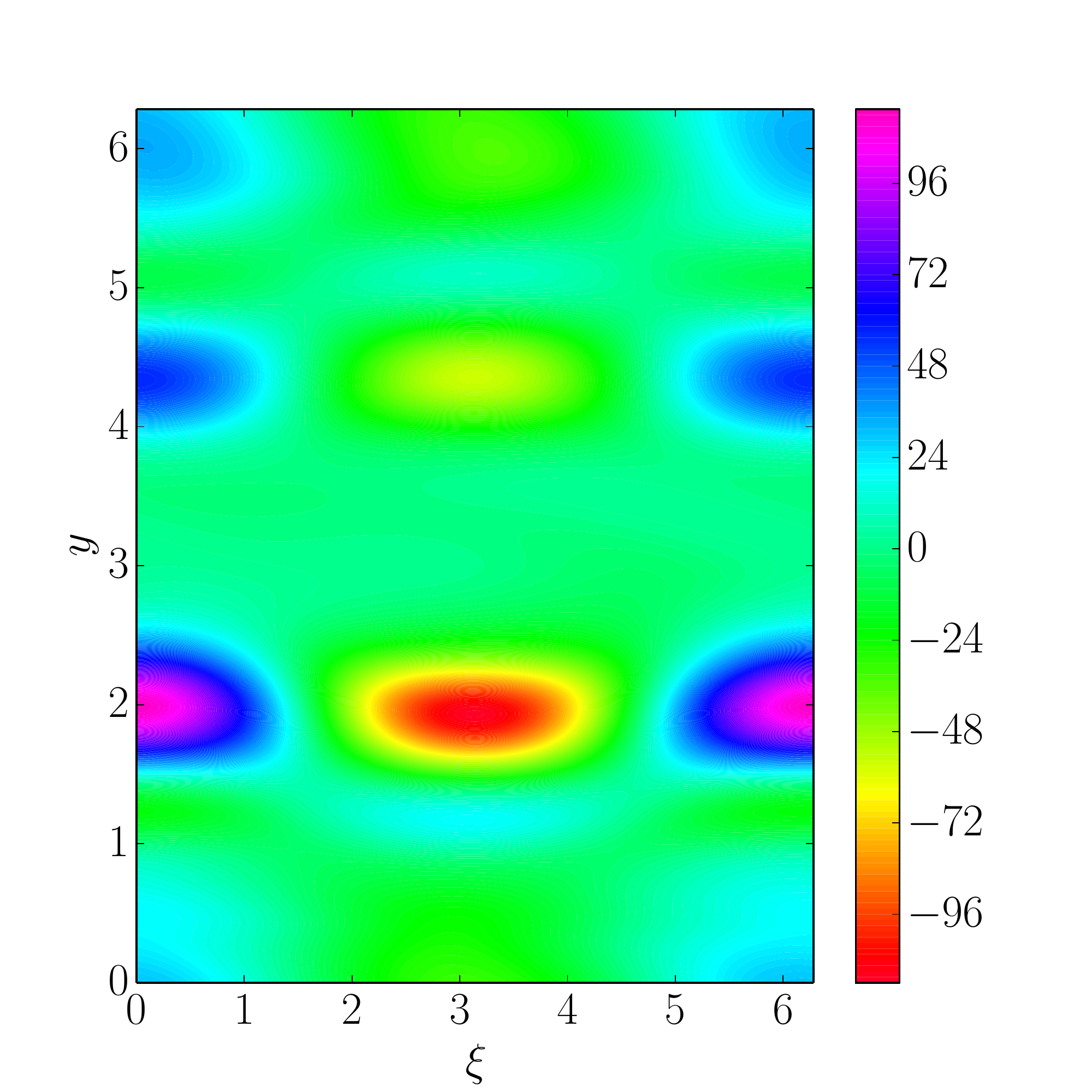} \quad \includegraphics[width=2.5in]{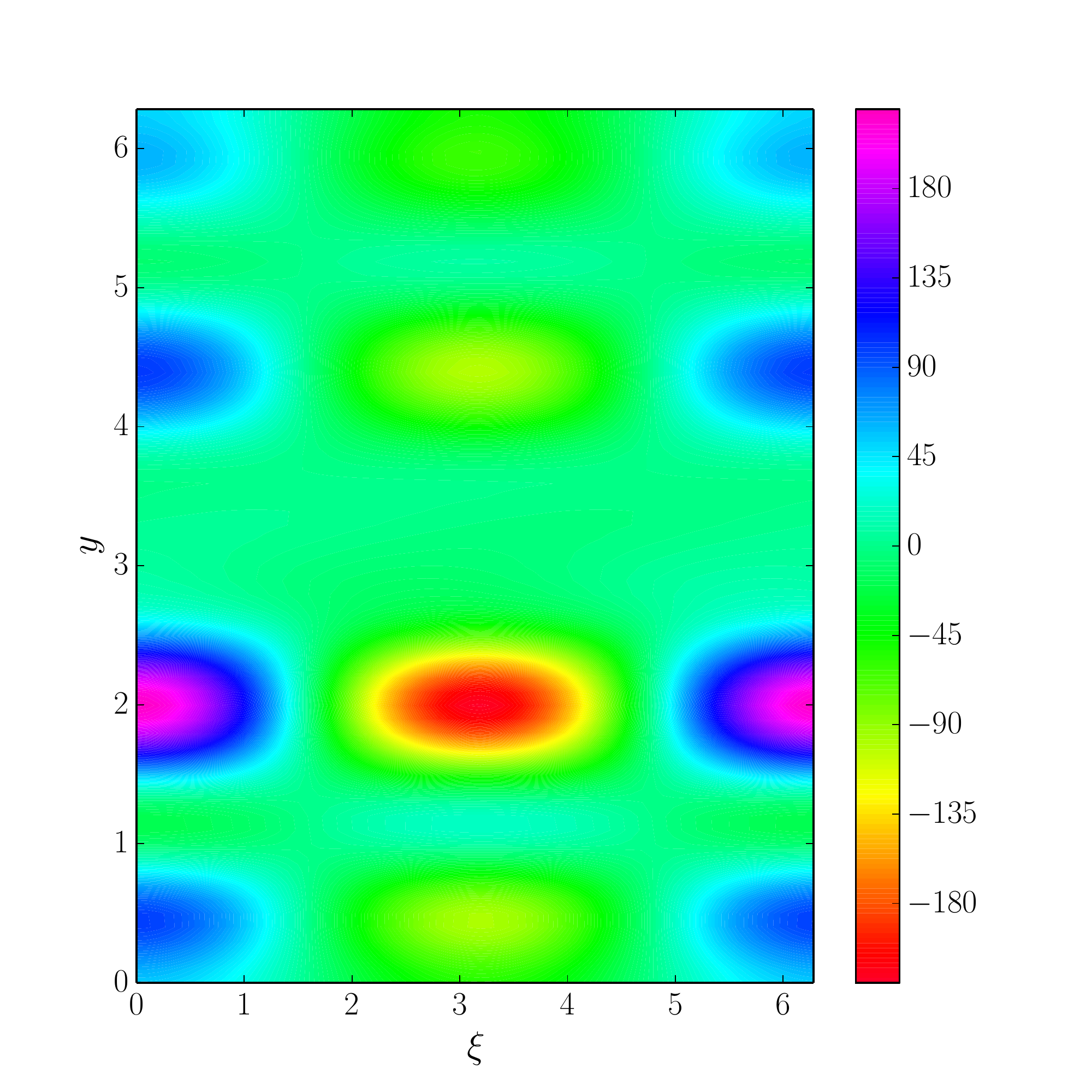}} 
\caption{$c_{\zeta \zeta}(y,3 \pi/4,\xi)$ for DNS (left) and CE2 (right) for the case $\nu=0.03$ (top row) $\nu=0.022$ (middle row) and $\nu =0.02$ (bottom row). In all cases $L_x=2 \pi$.}
\label{fig:sc_jets}
\end{figure*} 

\begin{figure*}
\centerline{\includegraphics[width=2.5in]{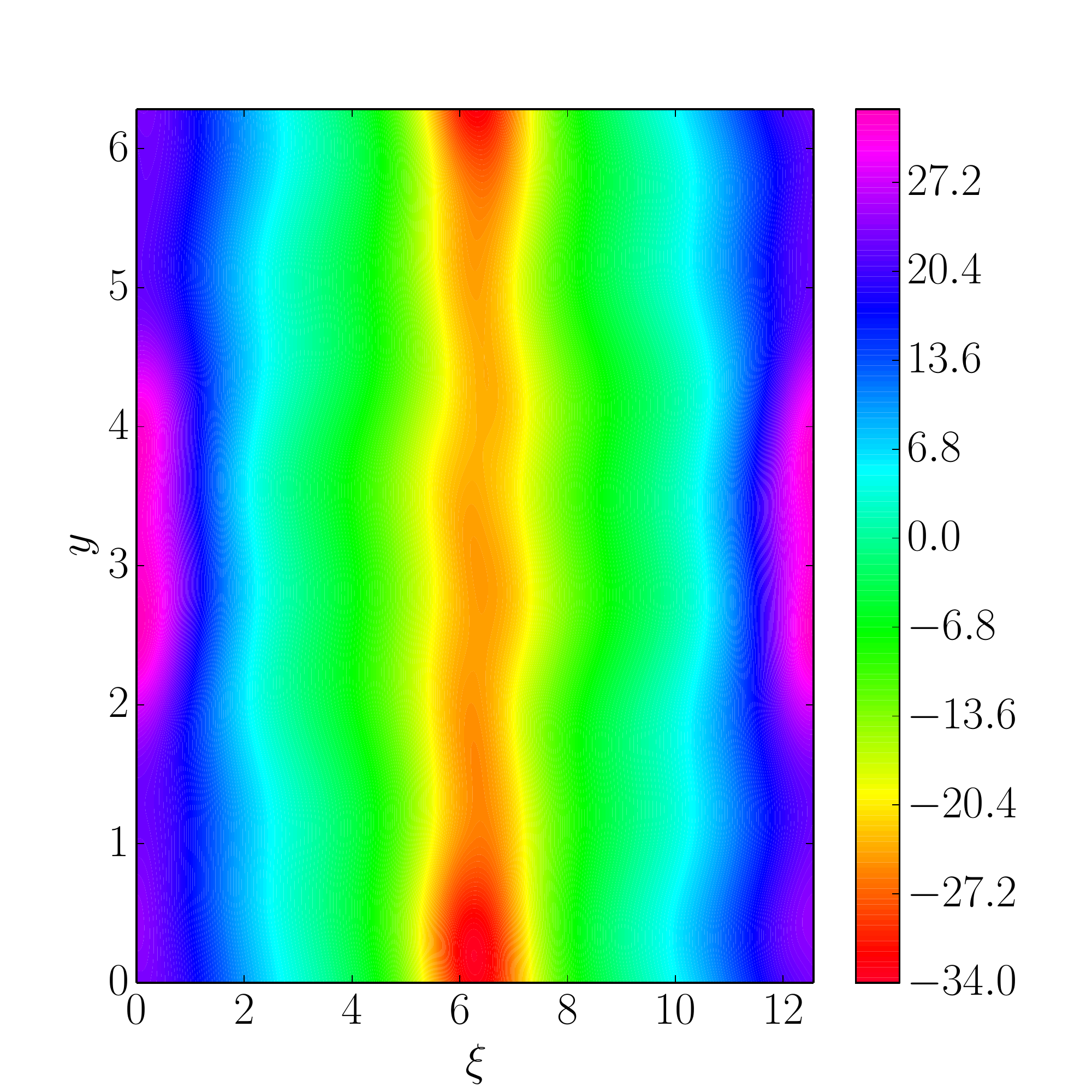} \quad \includegraphics[width=2.5in]{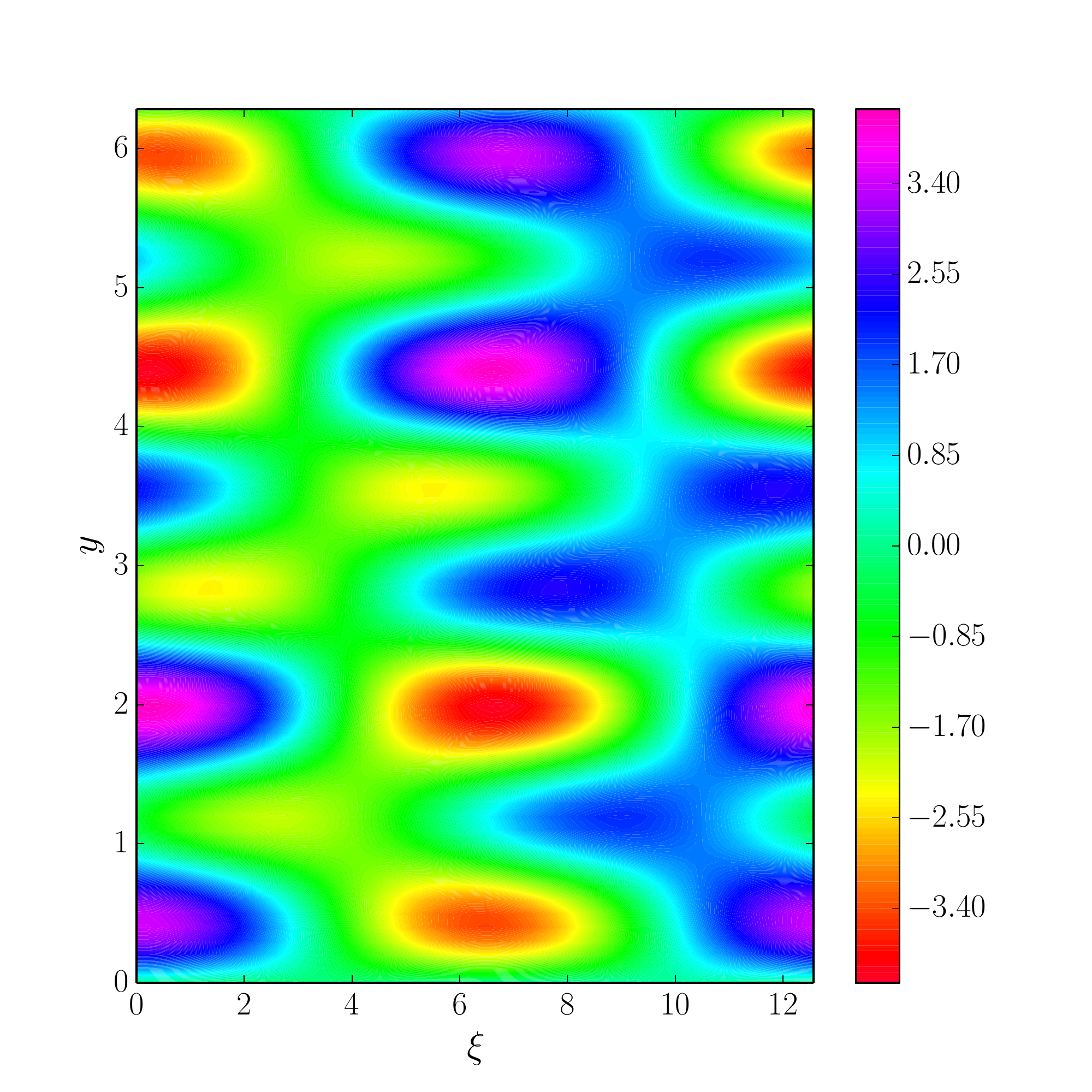}} 
\caption{$c_{\zeta \zeta}(y,3 \pi/4,\xi)$ for DNS (left) and CE2 (right) for the case $\nu=0.03$, $L_x=4 \pi$.}
\label{fig:sc_vort}
\end{figure*}

A more stringent test of the accuracy of statistical representation involves a comparison of not only the zonal means (first cumulants) but also the two-point correlation functions (second cumulants). These are given in Figures~\ref{fig:sc_jets} and \ref{fig:sc_vort}. These show the two-point correlation function $c_{\zeta \zeta}(y,3 \pi/4,\xi)$, i.e.\  the correlation in space  with the point three-eighths of the way up on the left hand side of each plot.  In all the cases for the jet solutions this is dominated by a $k_x=1, k_y=2$ solution, though this is modulated in $y$. DNS and CE2 can be seen to be in good agreement here as they should be for flows with such strong mean. Figure~\ref{fig:sc_vort} shows less good agreement between DNS and CE2, with CE2 failing to match both the amplitude and spatial form of the second cumulant (underestimating the $k_y=0$ component) for the case where the flow is dominated by strong vortices --- \smt{the DNS velocity correlation function has a near $k=0$ symmetry in the \jbm{meridional} direction and a near reflection symmetry in the zonal direction as the solution is dominated by a $k_x= 1$, $k_y=0$ vortex mode. This is clearly picked up by the correlation function, which is also dominated by these wavenumbers. The failure of CE2 to match the two-point correlation function for the vortex state} is not surprising as the form of this correlation function is presumably determined by eddy + eddy $\rightarrow$ eddy interactions that are discarded from the quasilinear CE2 description. 
\smt{The importance of cubic terms for the form of the solution is determined by the amplitude of their projection onto the second cumulant. This can be seen by comparing the second cumulants from CE2 and DNS. What is clear is that this projection is small for the cases with a significant zonal mean and large for the case of the vortex}
We discuss possible strategies for improving the agreement between DNS and DSS for this case \jbm{next} in the Discussion.

\section{Discussion}
\label{sec:conclusion}

We have shown that an expansion in equal-time \jbm{and zonally averaged} cumulants, truncated at second order, is able to describe both the jet- and vortex-dominated phases of a two-dimensional flow driven by deterministic Kolmogorov forcing.  Zonal mean flows in both phases are accurately reproduced, and two-point correlations of the vorticity are also captured in the jet phase, but not in the vortex-dominated phase.  

It is remarkable that DSS with such a simple closure can capture much of the behavior exhibited by DNS.  We expect that more sophisticated closures will be able to describe the vortex phase accurately.  Higher-order closures such as CE3$^*$ and CE2.5 (see Ref. \onlinecite{marston2014direct}) include eddy + eddy $\rightarrow$ eddy interactions and can improve qualitative agreement in two-point correlations found by DSS in comparison to DNS.  
The replacement of zonal averages with ensemble averages has been demonstrated to describe non-zonal structures \cite{Bakas:2015iy}, likely including the vortices seen here at aspect ratio $L_y / L_x = 2$.  Finally the generalized quasi-linear approximation (GQL)\cite{Marston:2016ff,childetal:2016} and its associated generalized cumulant expansion (GCE2), by allowing long-wavelength non-zonal structures to interact fully nonlinearly, should also be able to describe the vortex-dominated phase.  \jbm{Each of these variants is more computationally demanding than simple zonal-average CE2,} but we plan to test these other forms of DSS for the Kolmogorov forced model.  

All of these different forms of DSS respect the realizability inequalities studied by Kraichnan \cite{kraichnan1980realizability,Kraichnan:1985jy}.  They generalize the program of understanding the statistics of turbulence, greatly advanced by Kraichnan, to encompass anisotropy and heterogeneity.  Statistical theories of turbulence thus continue to extend their reach, permitting a deeper understanding of fluid flows that may someday allow us to access regimes not currently reachable by DNS.

\begin{acknowledgments}
We wish to acknowledge the help of Mark Dixon of The University of Leeds HPC facility team. All calculations were performed on the Arc2 machine at The University of Leeds.
\end{acknowledgments}


%

\end{document}